# Kinetic energy flows in activated dynamics of biomolecules

Huiyu Li and Ao Ma*

Department of Bioengineering

The University of Illinois at Chicago

851 South Morgan Street

Chicago, IL 60607

*correspondence should be addressed to:

Ao Ma

Email: aoma@uic.edu

Tel: (312) 996-7225

**Abstract**

Protein conformational changes are activated processes essential for protein functions. Activation in a protein differs from activation in a small molecule in that it involves directed and systematic energy flows through preferred channels encoded in the protein structure. Understanding the nature of these energy flow channels and how energy flows through them during activation is critical for understanding protein conformational changes. We recently (*J. Chem. Phys.*, **144**, 114103 (2016)) developed a rigorous statistical mechanical framework for understanding potential energy flows. Here we complete this theoretical framework with a rigorous theory for kinetic energy flows: potential and kinetic energy inter-convert when impressed forces oppose inertial forces whereas kinetic energy transfers directly from one coordinate to another when inertial forces oppose each other. This theory is applied to analyzing a prototypic system for biomolecular conformational dynamics: the isomerization of an alanine dipeptide. Among the two essential energy flow channels for this process, dihedral $\phi$ confronts the activation barrier, whereas dihedral $\theta_1$ receives energy from potential energy flows. Intriguingly, $\theta_1$ helps $\phi$ to cross the activation barrier by transferring to $\phi$ via direct kinetic energy flow all the energy it received—increase in $\dot{\theta}_1$ caused by potential energy flow converts into increase in $\dot{\phi}$. As a compensation, $\theta_1$ receives kinetic energy from bond angle $\alpha$ via direct mechanism and bond angle $\beta$ via indirect mechanism.

# I. Introduction

Proteins are the functional building blocks of cells and their conformational changes are essential to their functions. Understanding the mechanisms of protein conformational changes is critical to understanding protein functionality. A protein conformational change is an activated process: the protein molecule needs to cross an activation barrier significantly higher than thermal energy $k_B T$ [1, 2], where $k_B$ is the Boltzmann constant and $T$ is temperature.

In the standard physical picture of an activated process, the activation barrier is located on reaction coordinates, the slowest coordinates during activation; the system reaches the barrier top only during rare fluctuations [1, 3-12]. Accordingly, the numerous degrees of freedom (**DoF**s) in a complex molecular system (e.g. a protein molecule, a solution) are divided into reaction coordinates and heat bath. Reaction coordinates play a central role because they determine both the mechanism and the rate of activation. For example, to modify the activity of an enzyme, we should modify residues involved in the reaction coordinates for the enzymatic reaction [10, 13], as this will modify both the pathway and the barrier height for activation. In contrast, modifying residues that belong to the heat bath will not alter the enzymatic activity, as the role of heat bath is to provide energy to the reaction coordinates to cross the activation barrier during rare fluctuations, which is a non-specific process.

The importance of reaction coordinates had motivated search for rigorous methods to identify them in complex systems since early 2000's [4, 6, 14-17]. Beyond the intuition-based trial-and-error approach, the first systematic method was machine-learning, which used a neural network to automatically identify the optimal reaction coordinates from a pre-prepared pool of candidates [6]. This method was used to successfully identify the key solvent coordinate that controls the isomerization dynamics of an alanine dipeptide in solution, which had defied the intuition-based manual search. The success of this machine-learning approach inspired a series of developments along similar lines [15, 18-26].

However, a major deficiency of machine-learning methods is that they cannot answer the real question concerning reaction coordinates—why some coordinates are more important for activation than the others? Instead, they only inform us empirically which coordinates appear to be important based on well-defined criteria. Consequently, an approach based on first principles of physics (e.g. mechanical laws) is required to answer the real questions concerning reaction coordinates and the mechanism of activation, as activation dynamics are ultimately a consequence of the underlying mechanical laws.

We recently developed a rigorous theory for mapping out the flow of potential energy through individual coordinates [5, 27], defined as the work $\Delta W_i$ on a coordinate $q_i$. We applied this theory to the isomerization dynamics of an alanine dipeptide. This is a prototype for protein conformational dynamics because alanine dipeptide is the smallest molecule in which the non-reaction coordinates in the system can serve as a heat bath large enough to provide reaction coordinates with adequate energy to cross the activation barrier, the critical feature that distinguishes a complex molecule (e.g. a protein) from simple molecule (e.g. $CO_2$). We found that the reaction coordinates are the coordinates that carry high potential energy flows (**PEF**s) during activation. This result suggested an appealing physical picture: energy flows from fast coordinates into slow coordinates during activation so that adequate energy can accumulate in the slow coordinates to enable them to cross the activation barrier.

This physical picture also suggested that reaction coordinates are preferred channels of energy flows and they are encoded in the protein structure. In contrast, energy flows in small molecules are dominated by fast and largely homogeneous intramolecular vibrational energy redistribution that leads to quick equilibrium over the entire molecule [1, 2, 28-35]. Clearly, the energy flow analysis not only provides a rigorous way to identify reaction coordinates, but also provides invaluable mechanistic understanding that motivated the quest for reaction coordinates in the first place.

On the other hand, potential energy only constitutes half of the energy flows in a system—the other half is the kinetic energy flow (**KEF**). The importance of KEFs in understanding the mechanism of activation depends on the coordinate system in which activation dynamics are analyzed. In Cartesian coordinates, the work $\Delta W_\alpha$ on a given coordinate $x_\alpha$ completely converts into the change in its kinetic energy $K_\alpha = \frac{1}{2} m_\alpha \dot{x}_\alpha^2$. In this case, the KEF $\Delta K_\alpha$ contains identical information as the PEF $\Delta W_\alpha$ and does not provide any extra mechanistic information—there is no need to analyze KEFs in Cartesian coordinates.

However, Cartesian coordinates are not the natural coordinates for describing conformational dynamics of molecular systems. Potential energy flows in Cartesian coordinates will be dominated by work from strong and fast-fluctuating constraint forces that arise from bonded interactions and it is very challenging to deconvolute the true mechanistic information from the contamination from constraint forces. In contrast, internal coordinates (e.g. bond length and angle, dihedral) are the natural coordinates for describing protein conformational dynamics, as motions along these coordinates automatically satisfy the constraints from bonded interactions. Consequently, PEFs in internal coordinates contain clean mechanistic information that is ready for physical interpretation.

However, unlike Cartesian coordinates, internal coordinates are curvilinear. Consequently, the PEF $\Delta W_i$ into coordinate $q_i$ will not be converted into the kinetic energy of $q_i$, which is not well-defined because the system kinetic energy $K = \frac{1}{2}\sum_{i,j} s_{ij}\dot{q}_i\dot{q}_j$ contains cross terms $s_{ij}\dot{q}_i\dot{q}_j$ ($i \neq j$). Here, $\dot{q}_i$ is the velocity of coordinate $q_i$ and $s_{ij}$ is the structural coupling factor between coordinates $q_i$ and $q_j$. Instead, $\Delta W_i$ will spread into all the coordinates of the system, with the extent of spreading into any individual coordinate $q_j$ determined by the corresponding $s_{ij}$. Therefore, KEFs in internal coordinates contain important mechanistic information that is distinct from and complementary to the information in PEFs. A comprehensive mechanistic picture of conformational dynamics of proteins requires understanding both PEFs and KEFs.

In this paper, we present a general and rigorous theory for mapping and understanding KEFs during activation. We first introduce the theory and then demonstrate its usage by applying it to the isomerization of an alanine dipeptide in vacuum.

## II. Theory

In this section, we first review the theory for PEFs, before presenting the theory for KEFs.

**1. Brief review of potential energy flows**

The PEF through a given coordinate $q_i$ is its work [5]:

$$\Delta W_i(t_1, t_2) = \int_{q_i(t_1)}^{q_i(t_2)} F_i dq_i = -\int_{q_i(t_1)}^{q_i(t_2)} \frac{\partial U(\vec{q})}{\partial q_i} dq_i \quad (1)$$

Here, $F_i$ is the impressed force on $q_i$ (we call a force derived from potential energy an impressed force to distinguish it from an inertial force that is derived from kinetic energy) and $U(\vec{q})$ is the potential energy function of the system. According to this expression, $\Delta W_i(t_1, t_2)$ is the change in the potential energy of the system due to the motion of $q_i$ along a dynamic trajectory in the time interval $[t_1, t_2]$. It is a projection of the change in the total potential energy onto the motion of a specific coordinate. Therefore, it is a measure of the cost of the motion of a coordinate in terms of potential energy. Accordingly, the change in the total potential energy of the system can be decomposed into PEFs through different coordinates:

$$\Delta U(t_1, t_2) = U(t_2) - U(t_1) = -\sum_{i=1}^{N} \Delta W_i(t_1, t_2) \quad (2)$$

where the summation is over all coordinates of the system. A major finding from our previous PEF analysis was that reaction coordinates are the coordinates with high PEFs during activation.

**Proper average of per-coordinate PEF ($\Delta W_i$) to gain mechanistic information on activation**

To gain mechanistic insights into activation, we need to look at how the PEFs of individual coordinates change with the progress of activation. We first project the PEF onto a projector $\xi(\Gamma)$ that parameterizes the progress of activation, then average over the transition path ensemble (i.e. the ensemble of reactive trajectories):

$$\langle \delta W_i(\xi^*) \rangle = \frac{\int d\Gamma \rho(\Gamma) \delta W_i(\xi(\Gamma) \to \xi(\Gamma) + d\xi) \delta(\xi(\Gamma) - \xi^*)}{\int d\Gamma \rho(\Gamma) \delta(\xi(\Gamma) - \xi^*)}$$

$$\langle \Delta W_i(\xi_1 \to \xi_2) \rangle = \int_{\xi_1}^{\xi_2} \langle \delta W_i(\xi) \rangle \quad (3)$$

Here, $\rho(\Gamma)d\Gamma$ is the probability of finding the system in an infinitesimal volume $d\Gamma$ around a point $\Gamma$ in phase space in the transition path ensemble; $\delta(x)$ is the Dirac δ-function; $\delta W_i(\xi(\Gamma) \to \xi(\Gamma) + d\xi)$ is the change in $W_i$ in a differential interval $[\xi(\Gamma), \xi(\Gamma) + d\xi]$; $\langle \Delta W_i(\xi_1 \to \xi_2) \rangle$ is the change in $W_i$ in a finite interval $[\xi_1, \xi_2]$.

To understand the mechanism of an activated process, the projector $\xi(\Gamma)$ needs to properly parameterize the progress of activation. A straightforward choice is $\xi = p_B$, the so-called committor [4, 16, 36-38], defined as the probability that a dynamic trajectory initiated from a specific system configuration, with initial momenta drawn from Boltzmann distribution, reaches the product basin before the reactant basin. Committor is the reaction probability in configuration space; it provides a rigorous parameterization of the progress of an activated process.

**2. Kinetic energy flows**

Aside from PEFs, there are also KEFs. The mechanistic importance of KEFs in Cartesian coordinates and internal coordinates differs fundamentally.

In Cartesian coordinates, the work on a coordinate $x_\alpha$ (i.e. the PEF through $x_\alpha$) completely converts into the change in its kinetic energy:

$$dW_\alpha = -\frac{\partial U}{\partial x_\alpha} dx_\alpha = m_\alpha \ddot{x}_\alpha dx_\alpha = d\left(\frac{1}{2} m_\alpha \dot{x}_\alpha^2\right) = dK_\alpha \quad (4)$$

based on Newton's equation. Here, $K_\alpha = \frac{1}{2} m_\alpha \dot{x}_\alpha^2$ is the kinetic energy of $x_\alpha$ because it is a function of $\dot{x}_\alpha$ only. In this case, the KEF through $x_\alpha$ carry identical information as the PEF through $x_\alpha$ and there is no need to look into KEFs.

In contrast, there is no well-defined per-coordinate kinetic energy for internal coordinates because the system kinetic energy has cross-terms: $K = \frac{1}{2} \sum_{i,j} s_{ij} \dot{q}_i \dot{q}_j$. Here, $s_{ij} = \sum_\alpha m_\alpha \frac{\partial x_\alpha}{\partial q_i} \frac{\partial x_\alpha}{\partial q_j}$ is the structural coupling factor between coordinates $q_i$ and $q_j$; $s_{ij}$ is determined by the system structure. If we define the mass weighted Cartesian coordinate $X_\alpha = \sqrt{m_\alpha} x_\alpha$, then $s_{ij}$ can be rewritten more compactly: $s_{ij} = \sum_\alpha \frac{\partial X_\alpha}{\partial q_i} \frac{\partial X_\alpha}{\partial q_j} = \frac{\partial \vec{X}}{\partial q_i} \cdot \frac{\partial \vec{X}}{\partial q_j}$; $\vec{X}$ is the mass-weighted position vector of the entire system. Cartesian and internal coordinates have this fundamental difference because the former is straight whereas the latter is curvilinear.

Nevertheless, there is a well-defined KEF through a coordinate $q_i$:

$$\partial_v K_i = \frac{\partial K}{\partial \dot{q}_i} d\dot{q}_i + \left(\frac{\partial K}{\partial q_i}\right)_{\vec{q}',\vec{v}} dq_i = dt \left[p_i \ddot{q}_i + \left(\frac{\partial K}{\partial q_i}\right)_{\vec{q}',\vec{v}} \dot{q}_i\right] \quad (5),$$

where $\vec{q}' = (q_1, q_2, \cdots, q_{i-1}, q_{i+1}, \cdots, q_N)$ is the system position vector in internal coordinates with $q_i$ removed, and $\vec{v} = (\dot{q}_1, \dot{q}_2, \cdots, \dot{q}_N)$ is the velocity vector. Since $\partial_v K_i$ is the change in the system kinetic energy caused by changes in $(q_i, \dot{q}_i)$ alone, which fully describes the motion of $q_i$, it rigorously defines the KEF through $q_i$.

Unlike the case for Cartesian coordinates, the work on $q_i$ (i.e. the PEF through $q_i$) does not completely converts into the KEF through $q_i$. From Lagrange's equation [39]:

$$\frac{d}{dt}\left(\frac{\partial L}{\partial \dot{q}_i}\right)_{\vec{q},\vec{v}} = \dot{p}_i = -\left(\frac{\partial U}{\partial q_i}\right)_{\vec{q}'} + \left(\frac{\partial K}{\partial q_i}\right)_{\vec{q}',\vec{v}}, \qquad L(\vec{q},\dot{\vec{q}}) = K(\vec{q},\vec{v}) - U(\vec{q}) \quad (6),$$

we have:

$$dW_i = -\frac{\partial U}{\partial q_i} dq_i = \left[\dot{p}_i - \left(\frac{\partial K}{\partial q_i}\right)_{\vec{q}',\vec{v}}\right] dq_i = dt\left[\dot{p}_i \dot{q}_i - \left(\frac{\partial K}{\partial q_i}\right)_{\vec{q}',\vec{v}} \dot{q}_i\right] \neq \partial_v K_i \quad (7)$$

This means that the PEF through $q_i$ spreads into other coordinates in the system. Therefore, unlike in Cartesian coordinates, where there is a simple one-to-one inter-conversion between the PEF and KEF through a coordinate, the conversion between PEFs and KEFs in curvilinear coordinates is global and more complex.

The key to understand the interconversion between PEFs and KEFs in curvilinear coordinates is to realize that PEFs are work by impressed forces and KEFs are work by inertial forces [39]. Furthermore, energy flow is a result of forces opposing each other: potential energy converts into kinetic energy when an impressed force is opposed by an inertial force; kinetic energy transfers directly from one coordinate to another when two inertial forces oppose each other. The first situation is similar for Cartesian and curvilinear coordinates; the second situation is absent in Cartesian coordinates and unique to curvilinear coordinates. Therefore, the key to understand energy flows in curvilinear coordinates is to identify what forces are opposing each other and on which coordinate they act on.

We first clarify the essential facts of inertial forces. The inertial force due to the motions of a coordinate $q_i$, which we call the inertial force from $q_i$, is a vector of $N$ components: $\vec{\mathcal{F}}^i = (\mathcal{F}_1^i, \mathcal{F}_2^i, \ldots, \mathcal{F}_N^i)$. Each component acts on one of the $N$ coordinates in the system: $\mathcal{F}_j^i$ is the component of the inertial force from $q_i$ that acts on $q_j$. The sum of work by all these components is $\partial_v K_i = \sum_{j=1}^{N} \mathcal{F}_j^i dq_j = \vec{\mathcal{F}}^i \cdot d\vec{q}$—the KEF through $q_i$ is the total work by the inertial force from $q_i$. Similarly, on each coordinate $q_j$, there are $N$ inertial forces $(\mathcal{F}_j^1, \mathcal{F}_j^2, \ldots, \mathcal{F}_j^N)$ from the $N$ coordinates in the system. The sum of all these component forces, $\mathcal{F}_j = \sum_{i=1}^{N} \mathcal{F}_j^i$, is the total inertial force acting on $q_j$.

According to Lagrange's equation, the total impressed and inertial forces acting on $q_i$ oppose and exactly balance with each other: $F_i = \mathcal{F}_i$. Because $\mathcal{F}_i$ is a sum over $N$ components, each one from a different origin, this equation is actually a balance of $N + 1$ rather than two forces. On the other hand, we have the most straightforward physical picture when we can unambiguously identify a pair of forces opposing each other. Therefore, we want to identify clear pairwise relationships from the overall balance of $N + 1$ forces.

In the simplest case, the impressed force $F_i$ is of high magnitude and all the significant inertial forces are opposing it. In this case, the PEF through $q_i$ spreads into all coordinates–the amount that spreads into $q_j$ is $\mathcal{F}_i^j dq_i$. Alternatively, there are few strong inertial forces opposing each other—$\mathcal{F}_i$ is a result of cancelation between inertial forces. In this case, there are direct kinetic energy transfer between coordinates whose inertial forces oppose each other. For example, $\mathcal{F}_i^j dq_i > 0, \mathcal{F}_i^k dq_i < 0$ means kinetic energy directly flows from $q_k$ to $q_i$, because the source of kinetic energy increase in the former is the kinetic energy decrease in the latter, that is, $q_k$ is an energy donor whereas $q_j$ is an acceptor. On the other hand, if there are many forces of similar magnitudes belonging to two sets that oppose each other, it will be difficult to identify pairwise relationships. Instead, we can only draw a coarse-grained conclusion that energy flows from one set of coordinates to another.

Moreover, we need to identify the pairwise force-balance relationships in energy space instead of force space. This can be achieved by a closer examination of Hamilton's equation:

$$\dot{p}_i = -\left(\frac{\partial H}{\partial q_i}\right)_{\vec{q}',\vec{p}} = -\left(\frac{\partial U}{\partial q_i}\right)_{\vec{q}'} - \left(\frac{\partial K}{\partial q_i}\right)_{\vec{q}',\vec{p}}, \quad H(\vec{q},\vec{p}) = K(\vec{q},\vec{p}) + U(\vec{q}) \quad (8)$$

Comparing this with Eq. (6), we found $\left(\frac{\partial K}{\partial q_i}\right)_{\vec{q}',\vec{p}} = -\left(\frac{\partial K}{\partial q_i}\right)_{\vec{q}',\vec{v}}$ [39]. Consequently, both Lagrange's and Hamilton's equations mean:

$$dW_i = \left(\dot{p}_i + \left(\frac{\partial K}{\partial q_i}\right)_{\vec{q}',\vec{p}}\right)dq_i = \dot{q}_i dp_i + \left(\frac{\partial K}{\partial q_i}\right)_{\vec{q}',\vec{p}} dq_i = \frac{\partial K}{\partial p_i} dp_i + \left(\frac{\partial K}{\partial q_i}\right)_{\vec{q}',\vec{p}} dq_i = \partial_p K_i \quad (9)$$

From this equation, we can identify $\mathcal{F}_i = \sum_{j=1}^{N} \mathcal{F}_i^j = \dot{p}_i + \left(\frac{\partial K}{\partial q_i}\right)_{\vec{q}',\vec{p}}$ and $\partial_p K_i = \mathcal{F}_i dq_i$. This equation suggests that the potential energy flowing into $q_i$ completely converts into $\partial_p K_i$, which is associated with $(q_i, p_i)$, instead of $\partial_v K_i$, which is associated with $(q_i, \dot{q}_i)$. Because $p_i = \sum_k s_{ik} \dot{q}_k$ is a many-body quantity that depends on all DoFs of the system, $\partial_p K_i$ is not the KEF through $q_i$ or any specific coordinate.

Since $\partial_p K_i$ contains the work by all the inertial forces (i.e. $\mathcal{F}_i^k$) acting on $q_i$ and $\partial_v K_j$ contains all the work by the inertial force from $q_j$, the term common to $\partial_p K_i$ and $\partial_v K_j$, which we call $c_{ip,jv}$, is the work by the inertial force from $q_j$ that acts on $q_i$—$c_{ip,jv} = \mathcal{F}_i^j dq_i$. Therefore, identifying $c_{ip,jv}$ for $j = 1, \dots, N$ allows us to identify all the inertial forces acting on $q_i$ and determine which forces oppose each other based on the signs of different work contributions. If $sign(dW_i) = sign(\mathcal{F}_i^j dq_i)$, then $\mathcal{F}_i^j$ opposes $F_i$; if $sign(dW_i) = -sign(\mathcal{F}_i^j dq_i)$ instead, then $\mathcal{F}_i^j$ and $F_i$ act in the same direction. Similarly, if $sign(\mathcal{F}_i^j dq_i) = -sign(\mathcal{F}_i^k dq_i)$, then $\mathcal{F}_i^j$ opposes $\mathcal{F}_i^k$; otherwise, they act in the same direction.

To identify $c_{ip,jv}$, we need to expand $\partial_p K_i$ and $\partial_v K_j$ into detailed terms. We start from $\dot{p}_i = \sum_k (\dot{s}_{ik} \dot{q}_k + s_{ik} \ddot{q}_k)$, which gives:

$$\dot{p}_i \dot{q}_i = \sum_k (\dot{s}_{ik} \dot{q}_i \dot{q}_k + s_{ik} \dot{q}_i \ddot{q}_k) = \sum_{k,l} \frac{\partial s_{ik}}{\partial q_l} \dot{q}_i \dot{q}_k \dot{q}_l + \sum_k s_{ik} \dot{q}_i \ddot{q}_k$$

$$= \sum_{k,l} \dot{q}_i \dot{q}_k \dot{q}_l \left[\sum_\alpha \left(\frac{\partial^2 X_\alpha}{\partial q_i \partial q_l} \frac{\partial X_\alpha}{\partial q_k} + \frac{\partial^2 X_\alpha}{\partial q_k \partial q_l} \frac{\partial X_\alpha}{\partial q_i}\right)\right] + \sum_k \dot{q}_i \ddot{q}_k \sum_\alpha \frac{\partial X_\alpha}{\partial q_i} \frac{\partial X_\alpha}{\partial q_k} \quad (10).$$

In addition, we have:

$$\left(\frac{\partial K}{\partial q_i}\right)_{\vec{v}} \dot{q}_i = \frac{1}{2} \sum_{k,l} \frac{\partial s_{kl}}{\partial q_i} \dot{q}_i \dot{q}_k \dot{q}_l = \frac{1}{2} \sum_{k,l} \left[\sum_\alpha \left(\frac{\partial^2 X_\alpha}{\partial q_i \partial q_k} \frac{\partial X_\alpha}{\partial q_l} + \frac{\partial^2 X_\alpha}{\partial q_i \partial q_l} \frac{\partial X_\alpha}{\partial q_k}\right)\right] \dot{q}_i \dot{q}_k \dot{q}_l$$

$$= \dot{q}_i \sum_{k,l} \dot{q}_k \dot{q}_l \left( \sum_\alpha \frac{\partial^2 X_\alpha}{\partial q_i \partial q_l} \frac{\partial X_\alpha}{\partial q_k} \right) = \dot{q}_i \sum_\alpha \sum_k \frac{\partial X_\alpha}{\partial q_k} \dot{q}_k \sum_l \frac{\partial^2 X_\alpha}{\partial q_i \partial q_l} \dot{q}_l$$

$$= \dot{q}_i \sum_\alpha \sum_k \frac{\partial X_\alpha}{\partial q_k} \dot{q}_k \frac{\partial}{\partial q_i} \left( \sum_l \frac{\partial X_\alpha}{\partial q_l} \dot{q}_l \right) = \dot{q}_i \frac{\partial \dot{\vec{X}}}{\partial q_i} \cdot \left( \sum_k \frac{\partial \vec{X}}{\partial q_k} \dot{q}_k \right)$$

$$= \dot{q}_i \frac{\partial \dot{\vec{X}}}{\partial q_i} \cdot \frac{\partial \vec{X}}{\partial \vec{q}} \cdot \dot{\vec{q}} \quad (11).$$

Here, $\dot{\vec{X}}$ is the time derivative of the position vector $\vec{X}$. Consequently, we have:

$$\partial_p K_i = dt \left[ \dot{p}_i \dot{q}_i - \left( \frac{\partial K}{\partial q_i} \right)_{\vec{v}} \dot{q}_i \right] = dt \left[ \dot{q}_i \sum_{k,l} \dot{q}_k \dot{q}_l \left( \sum_\alpha \frac{\partial^2 X_\alpha}{\partial q_k \partial q_l} \frac{\partial X_\alpha}{\partial q_i} \right) + \dot{q}_i \sum_k s_{ik} \ddot{q}_k \right]$$

$$= dq_i \sum_k \frac{\partial \vec{X}}{\partial q_i} \cdot \left( \frac{\partial \dot{\vec{X}}}{\partial q_k} \dot{q}_k + \frac{\partial \vec{X}}{\partial q_k} \ddot{q}_k \right) = \frac{\partial \vec{X}}{\partial q_i} \cdot \left( \frac{\partial \dot{\vec{X}}}{\partial \vec{q}} \cdot \dot{\vec{q}} + \frac{\partial \vec{X}}{\partial \vec{q}} \cdot \ddot{\vec{q}} \right) dq_i \quad (12)$$

On the other hand, we have:

$$\partial_v K_j = dt \left[ p_j \ddot{q}_j + \left( \frac{\partial K}{\partial q_i} \right)_{\vec{v}} \dot{q}_j \right] = dt \left[ \dot{q}_j \sum_{k,l} \dot{q}_k \dot{q}_l \left( \sum_\alpha \frac{\partial^2 X_\alpha}{\partial q_j \partial q_l} \frac{\partial X_\alpha}{\partial q_k} \right) + \ddot{q}_j \sum_k s_{jk} \dot{q}_k \right]$$

$$= dt \left[ \sum_k \dot{q}_k \frac{\partial \vec{X}}{\partial q_k} \cdot \left( \frac{\partial \vec{X}}{\partial q_j} \ddot{q}_j + \frac{\partial \dot{\vec{X}}}{\partial q_j} \dot{q}_j \right) \right] = \left( \ddot{q}_j \frac{\partial \vec{X}}{\partial q_j} + \dot{q}_j \frac{\partial \dot{\vec{X}}}{\partial q_j} \right) \cdot \frac{\partial \vec{X}}{\partial \vec{q}} \cdot d\vec{q} \quad (13).$$

Therefore, terms shared by $\partial_p K_i$ and $\partial_v K_j$ are:

$$c_{ip,jv} = s_{ij} \dot{q}_i \ddot{q}_j + \dot{q}_i \dot{q}_j \sum_l \dot{q}_l \left( \sum_\alpha \frac{\partial^2 X_\alpha}{\partial q_j \partial q_l} \frac{\partial X_\alpha}{\partial q_i} \right) = \frac{\partial \vec{X}}{\partial q_i} \cdot \left( \frac{\partial \vec{X}}{\partial q_j} \ddot{q}_j + \frac{\partial \dot{\vec{X}}}{\partial q_j} \dot{q}_j \right) dq_i \quad (14).$$

In contrast, $\partial_p K_i$ and $\partial_p K_j$ have no terms in common; $\partial_v K_i$ and $\partial_v K_j$ have no terms in common. We also have rigorous decompositions of both $\partial_p K_i = \sum_j c_{ip,jv}$ and $\partial_v K_i = \sum_j c_{jp,iv}$.

We further note that the inertial force $\mathcal{F}_i^j$ has two terms. The term $\mathcal{F}_i^{j,a} = \frac{\partial \vec{X}}{\partial q_i} \cdot \frac{\partial \vec{X}}{\partial q_j} \ddot{q}_j$ is due to the acceleration of $q_j$ and is similar to its counterpart in Cartesian coordinate, $m_\alpha \ddot{x}_\alpha$, that we are familiar with. We call $\mathcal{F}_i^{j,a}$ the acceleration force from $q_j$ that acts on $q_i$. The other term, $\mathcal{F}_i^{j,r} = \frac{\partial \vec{X}}{\partial q_i} \cdot \frac{\partial \dot{\vec{X}}}{\partial q_j} \dot{q}_j$ has no counterpart in Cartesian coordinates. It is due to the coupling between different coordinates and is responsible for redistributing kinetic energy among different coordinates. We call $\mathcal{F}_i^{j,r}$ the redistribution force from $q_j$ that acts on $q_i$. This term will not vanish even when there is no impressed force in a system.

## 3. An example: KEF analysis on a particle moving in a central force field

To illustrate the use of KEF analysis, we apply it to the familiar example of a particle moving in a central force field: its angular velocity increases when it moves towards the center of the force field even though there is no impressed force acting on the angular coordinate (Fig. 1). Another example that follows the same mechanism is the spin move of a skater: a skater can speed up her spin by pulling her arms and leg towards her body. With the conventional approach, we can infer that energy must flow from the radial coordinate $r$ to the angular coordinate $\theta$ based on energy conservation considerations, but we do not know how this happens. The energy flow formalism, in contrast, enables us to determine precisely how energy flows from $r$ to $\theta$ and the specific roles played by the impressed and inertial forces in this process.

Based on the general formalism discussed in previous sections, on $r$ coordinate we have:
$$\partial_p K_r = c_{rp,rv} + c_{rp,\theta v} = (\mathcal{F}_r^{r,a} + \mathcal{F}_r^{\theta,r})dr = m\ddot{r}dr - mr\dot{\theta}^2 dr = dW_r > 0 \quad (15)$$
The two inertial forces, $\mathcal{F}_r^{r,a} = m\ddot{r}$ and $\mathcal{F}_r^{\theta,r} = -mr\dot{\theta}^2$, both oppose the central force $F_r$, whereas $\mathcal{F}_r^{r,r} = \mathcal{F}_r^{\theta,a} = 0$. While $c_{rp,rv}$ is the work by $\mathcal{F}_r^{r,a}$, the centrifugal force, and converts $dW_r$ into KEF in $r$, $c_{rp,\varphi v}$ is the work by $\mathcal{F}_r^{\theta,r}$ and converts $dW_r$ into KEF in $\theta$. The net result is that $\mathcal{F}_r^{\theta,r}$ converts a fraction of the PEF in $r$ into KEF in $\theta$.

With regard to forces on $\theta$, we have:
$$\partial_p K_\theta = c_{\theta p,rv} + c_{\theta p,\theta v} = \left(\mathcal{F}_\theta^{r,r} + \left(\mathcal{F}_\theta^{\theta,r} + \mathcal{F}_\theta^{\theta,a}\right)\right)d\theta$$
$$= \left(mr\dot{r}\dot{\theta} + (mr\dot{r}\dot{\theta} + mr^2\ddot{\theta})\right)d\theta = dW_\theta = 0 \quad (16)$$
In this case, $c_{\theta p,rv} = mr\dot{r}\dot{\theta}d\theta = mr\dot{\theta}^2 dr = -c_{rp,\theta v} < 0$, thus we have $c_{\theta p,\theta v} = mr\dot{r}\dot{\theta}d\theta + mr^2\ddot{\theta}d\theta > 0$, which means $mr^2\ddot{\theta}d\theta > 0$. Since there is no impressed force on $\theta$, the acceleration force from $\theta$ ($\mathcal{F}_\theta^{\theta,a} = mr^2\ddot{\theta}$) is balanced by the redistribution forces from both $r$ ($\mathcal{F}_\theta^{r,r} = mr\dot{r}\dot{\theta}$) and $\theta$ ($\mathcal{F}_\theta^{\theta,r} = mr\dot{r}\dot{\theta}$). The kinetic energy extracted from $r$ by $c_{\theta p,rv}$ became the kinetic energy increase in $\theta$ due to $c_{\theta p,\theta v}$. Therefore, it is a direct transfer of kinetic energy from $r$ to $\theta$. This is the second mechanism for energy transfer between $r$ and $\theta$.

In both mechanisms discussed above, $r$ is the energy donor and $\theta$ is the acceptor. Altogether, the total amount of kinetic energy transferred from $r$ to $\theta$ via these two mechanisms is $c_{rp,\theta v} + c_{\theta p,\theta v} = c_{rp,\theta v} - c_{\theta p,rv} = -mr\dot{\theta}^2 dr - mr\dot{r}\dot{\theta}d\theta = -2mr\dot{\theta}^2 dr = mr^2\ddot{\theta}d\theta$.

**III. Simulation Results**

In this section, we apply the theory for KEF analysis to the $C_{7eq} \to C_{7ax}$ transition of an alanine dipeptide in vacuum [5, 6, 14], which is mainly the rotation around the $2N - 2C_\alpha$ bond that defines the $\phi$ dihedral angle (Fig. 2). We analyzed the PEFs of this process in a previous study [5], and found that the dominant reaction coordinates are $\phi$ and $\theta_1$, which carry high PEFs during the isomerization process. This is because the PEFs are closely related to the time scales of coordinates—energy flows from fast into slow coordinates. Moreover, three additional coordinates, dihedral $\psi$ and bond angles $\alpha$ and $\beta$, have significant exchange of potential energy with $\phi$ and $\theta_1$, even though they do not have high PEFs through them. This makes them also important for the isomerization dynamics. On the other hand, the theory for KEF analysis used in that study was inadequate.

The quantity we used for KEF analysis in the previous study was the kinetic virial $K_i = \frac{1}{2}p_i\dot{q}_i$. By comparing with Eqs. (5) and (9), we have: $\frac{1}{2}(\partial_p K_i + \partial_v K_i) = \frac{1}{2}(\dot{q}_i dp_i + p_i d\dot{q}_i) = dK_i$. Because $dW_i = \partial_p K_i$, we have $dW_i - dK_i = \frac{1}{2}(dW_i - \partial_i K_v)$. Therefore, $\langle \Delta W_i \rangle - \langle \Delta K_i \rangle$ reflects the amount of PEF in $q_i$ that leaks into other coordinates. Although this quantity cannot provide mechanistic information on KEFs, it provides an approximate overall count.

The major conclusion from the KEF analysis using kinetic virial was that $\theta_1$ transfers a fraction of the PEF it received from the other coordinates of the system into the KEF in $\phi$ to help $\phi$ cross the activation barrier, which is located on the path of $\phi$ motion. This was an intriguing result, but it was marred by uncertainty due to the inadequacy of the KEF analysis with kinetic virial. With a rigorous theory for KEF analysis at hand, we now revisit the KEFs in this process.

We start with examining $\Delta_v K_i$ for all the DoFs in the system. Figure 3 shows that non-vanishing KEFs only exist for the five important coordinates, $(\phi, \theta_1, \psi, \alpha, \beta)$, identified by the PEF analysis. Moreover, $\langle \Delta K_{tot} \rangle \simeq \langle \Delta_v K_\phi \rangle + \langle \Delta_v K_{\theta_1} \rangle + \langle \Delta_v K_\psi \rangle + \langle \Delta_v K_\alpha \rangle + \langle \Delta_v K_\beta \rangle$, confirming that KEFs through these five coordinates account for the total KEF through the system.

In addition, Fig. 4 shows that $|\Delta_v K_\phi| < |\Delta W_\phi|, |\Delta_v K_{\theta_1}| < |\Delta W_{\theta_1}|$. This result suggests that a significant fraction of the PEF into $\theta_1$ leaks into other coordinates, and $\phi$ receives significant KEFs from other coordinates to help it cross the activation barrier. In addition, we have $|\Delta W_\phi + \Delta W_{\theta_1}| \simeq |\Delta_v K_\phi|$, suggesting that the KEF into $\phi$ might be from $\theta_1$.

To identify the KEFs through $\phi$, we examined $c_{ip,\phi v}$ with $i = 1, \ldots, N$ covering all the coordinates in the system. Figure 5 showed that $c_{ip,\phi v}$ is significant for $i = \phi, \theta_1, \psi, \beta, \gamma, \omega, \tau$, suggesting that $\phi$ exchanged significant kinetic energy with other coordinates through $\mathcal{F}_i^\phi$ on these coordinates. To identify the donors and acceptors in the kinetic energy exchanges in these cases, we examined $c_{ip,jv}$ with $i = \phi, \theta_1, \psi, \beta, \gamma, \omega, \tau$ and $j = 1, \ldots, N$. We found that $\psi, \alpha, \beta$ acted as both donor and acceptor in their kinetic energy exchanges with $\phi$, whereas $\theta_1$ acted only as donor.

The kinetic energy exchange between $\theta_1$ and $\phi$ is due to their inertial forces that oppose each other on $\tau$, the improper dihedral that maintains the planar configuration of the carbonyl group that is involved in $\theta_1$. Figure 6 showed that $c_{\tau p,\phi v} \approx \Delta W_{\theta_1}$, suggesting that the kinetic energy that $\phi$ received from $\theta_1$ equals to the energy that $\theta_1$ received from PEFs. In this case, the inertial force from $\theta_1$ is the only force that oppose the inertial force from $\phi$, thus we can unambiguously identify $\theta_1$ as the kinetic energy donor to $\phi$. Therefore, the data suggest that $\theta_1$ converted all the energy it gained from PEFs into KEF into $\phi$ to help $\phi$ cross the activation barrier.

In addition, the inertial forces from $\theta_1$ and $\phi$ that act on $\tau$ are mostly acceleration forces, which are directly related to changes in $\dot\theta_1$ and $\dot\phi$. This means that decrease in $\dot\theta_1$ directly converts into increase in $\dot\phi$. Moreover, Fig. 7 showed that $c_{\theta_1 p,\theta_1 v}$ is mainly due to acceleration force of $\theta_1$ as well, and it is opposed by the impressed force on $\theta_1$, suggesting the PEF into $\theta_1$ directly increased $\dot{\theta}_1$. Therefore, the impressed force on $\theta_1$ indirectly accelerated the motion of $\phi$, which is achieved by directly transferring kinetic energy from $\theta_1$ to $\phi$. That is, the impressed force on $\theta_1$ translated into the acceleration force of $\theta_1$ acting on $\tau$, which opposes the acceleration force of $\phi$ acting on $\tau$ and increases $\dot{\phi}$. Moreover, we have $c_{\tau p, \phi v} \simeq |\Delta W_\phi - \Delta_v K_\phi|$, confirming that all the kinetic energy gained by $\phi$ is from $\theta_1$.

In contrast, the way in which $\phi, \psi, \beta, \gamma, \omega, \tau$ exchange kinetic energy with $\phi$ is not one-to-one. Instead, it always happens that one set of coordinates donate kinetic energy to another set of coordinates. As discussed above, we cannot uniquely determine the amount of kinetic energy exchange between any pair of coordinates under this situation. Consequently, the kinetic energy exchanges can be assigned such that the net exchange between $\phi$ and any of these coordinates is zero, which is consistent with the fact that the unidirectional flow of kinetic energy from $\theta_1$ to $\phi$ can account for all the net gain in kinetic energy by $\phi$. Therefore, the significant energy flows between $\phi$ and $\psi$, $\beta, \gamma, \omega, \tau$ without net exchange mainly reflect the strong coupling between the motions of these coordinates. This is likely the reason that conformational changes in complex molecules often requires global and collective motion of many local coordinates.

Since $\theta_1$ transferred all the energy it gained from PEFs to $\phi$ via direct transfer of kinetic energy, we expect $\langle \Delta_v K_{\theta_1} \rangle \simeq 0$. However, $\langle \Delta_v K_{\theta_1} \rangle$ is significant, suggesting that $\theta_1$ gained kinetic energy from other coordinates. Since $\theta_1$ can only have net gain in kinetic energy from coordinates that suffer net loss in kinetic energy, the only candidates are $\alpha$ and $\beta$. Indeed, Fig. 3 shows that $\langle \Delta_v K_{\theta_1} \rangle + \langle \Delta_v K_\alpha \rangle + \langle \Delta_v K_\beta \rangle \simeq 0$, suggesting that the sources for the net KEFs into $\theta_1$ are likely $\alpha$ and $\beta$, though such transfers can be either direct or mediated by other coordinates.

To answer this question, we examined $c_{ip, \theta_1 v}$ with $i = 1, ..., N$. Figure S1 shows that $\langle c_{\beta p, \theta_1 v} \rangle \simeq \langle \Delta_v K_{\theta_1} \rangle$, suggesting the net KEFs into $\theta_1$ occurred on $\beta$. To identify the donors of kinetic energy to $\theta_1$, we examined all the components of $\Delta_p K_\beta$. Figure S1 showed that $\alpha, \gamma, \omega$ are the potential donors. From this, we conclude that $\alpha$ transferred kinetic energy to $\theta_1$ via the direct mechanism.

On the other hand, $\langle \Delta_v K_\gamma \rangle \simeq 0$ and $\langle \Delta_v K_\omega \rangle \simeq 0$, suggesting that they cannot provide net KEF to $\theta_1$. Instead, they can only mediate indirect transfer of kinetic energy to $\theta_1$ from other coordinates––if they transferred kinetic energy to $\theta_1$ on $\beta$, they must receive kinetic energy from other coordinates somewhere else to compensate this loss in kinetic energy.

To answer this question, we examined $c_{ip,\gamma v}$ and $c_{ip,\omega v}$ with $i = 1, ..., N$. Figure S2 showed that $\gamma$ gained significant kinetic energy on $\alpha$, thus we examined all the components of $\Delta_p K_\alpha$, which showed that $\gamma$ gained kinetic energy from $\beta$ on $\alpha$. This result suggests that kinetic energy transferred from $\beta$ to $\gamma$ and then from $\gamma$ to $\theta_1$—$\gamma$ mediated kinetic energy transfer from $\beta$ to $\theta_1$ via indirect mechanism. Similarly, Fig. S3 showed that $\omega$ gained kinetic energy on $\psi$, thus we examined all the components of $\Delta_p K_\psi$, which showed that $\omega$ gained kinetic energy from $\beta$, again suggests a indirect kinetic energy transfer from $\beta$ to $\theta_1$. Together, these results showed net KEFs into $\theta_1$ from $\alpha$ and $\beta$.

In summary, the PEF analysis revealed that the activation barrier is located on the path of $\phi$ motion, whereas $\theta_1$ actually receives significant energy from other coordinates in the system. However, $\theta_1$ donates all the energy it received from PEFs to $\phi$ by directly transferring kinetic energy to $\phi$, which is achieved via the balance between the acceleration forces of $\theta_1$ and $\phi$ acting on $\tau$, so that the decrease in $\dot{\theta}_1$ directly converts into increase in $\dot{\phi}$ to help $\phi$ cross the activation barrier. In addition, $\theta_1$ receives kinetic energy from $\alpha$ via direct transfer and from $\beta$ via indirect mechanism mediated by $\gamma$ and $\omega$. Finally, there are significant KEFs between the five important coordinates $\phi, \theta_1, \psi, \alpha, \beta$, though with no net kinetic energy exchange, reflecting the strong coupling between their motions that leads to the collective behavior usually observed in the conformational dynamics of complex molecules. Without the rigorous theory on KEF analysis, it would have been impossible to obtain these precise mechanistic insights.

**IV. Discussions**

In this paper, we presented a rigorous theory for KEFs in complex molecules. Together with the theory for PEFs, they form a comprehensive theoretical framework for understanding energy flows during activations in complex molecular systems. Based on the analysis of PEFs and KEFs in the isomerization dynamics of an alanine dipeptide in vacuum, we can infer some general conclusions regarding energy flows in activated processes of biomolecules, which could have important implications in applications such as enhanced sampling [40-42].

Based on the features of PEFs and KEFs, the coordinates in a system can be classified into four categories. The first category consists of coordinates with high PEFs and high KEFs (e.g. $\phi$ and $\theta_1$ in the alanine dipeptide example). These coordinates are the essential reaction coordinates. The second category consists of coordinates with appreciable per-coordinate PEFs or KEFs and high coordinate-to-coordinate PEFs and KEFs with the essential coordinates and among themselves (e.g. $\psi, \alpha, \beta$). These coordinates are likely part of reaction coordinates, though they play a minor role compared to coordinates in the first category. The third category consists of coordinates with neither per-coordinate PEFs or KEFs themselves, nor do they have significant coordinate-to-coordinate PEFs or KEFs with coordinates in the first two categories. However, they act as mediators for significant coordinate-to-coordinate kinetic energy exchanges between coordinates in the first two categories (e.g. $\gamma, \tau, \omega$). The final category contains coordinates that has no high magnitude energy flows of any kind. They form the heat bath.

Another important observation is that the PEFs and KEFs among the important coordinates (e.g. $\phi, \theta_1, \psi, \alpha, \beta$) show a complex network structure. This is likely the reason that protein activation dynamics in general involve global and collective motions. Rigorous method for analyzing network topology is required to uncover the connection between the energy flow network and the collectivity in protein activation dynamics.

The curvilinear nature of internal coordinates leads to different mechanisms for KEFs. One mechanism is the direct conversion of the PEF of one coordinate $q_i$ into the KEF of a different coordinate $q_j$. This happens when the inertial force from $q_j$ opposes the impressed force on $q_i$. Another mechanism is the direct exchange of kinetic energy between two different coordinates when their inertial forces directly oppose each other. In the alanine dipeptide example, direct KEF from $\theta_1$ to $\phi$ happens when their inertial forces both act on $\tau$ and oppose each other. Finally, it is also possible that kinetic energy flows indirectly from $q_i$ to $q_j$ via the mediation of $q_k$. In this case, there will be direct KEFs from $q_i$ to $q_k$, and then from $q_k$ and $q_j$. These two coordinate-to-coordinate KEFs should be of opposite sign and equal magnitude, so that overall there is no net KEF in $q_k$, but there is significant net KEF from $q_i$ to $q_j$.

**Simulation Details**

All simulations were performed using the molecular dynamics software suite GROMACS-4.5.4 [43] with transition path sampling implemented. Amber 94 force field was used for consistency with previous results [44-46]. The structure of the alanine dipeptide was minimized using steepest descent algorithm and heated to 300 K using velocity rescaling with a coupling constant of 0.2 ps [47]. The system was then equilibrated for 200 ps and no constraints were applied. The time step of integration was 1 fs. Basin $C_{7eq}$ was defined as $-200° < \phi < -55°$ and $-90° < \psi < 190°$; basin $C_{7ax}$ was defined as $50° < \phi < 100°$ and $-80° < \psi < 0°$. Transition path sampling was used to harvest 3,5000 independent reactive trajectories from $C_{7eq}$ to $C_{7ax}$. Transition paths were 2 ps in length and simulated with a constant energy of 36 kJ/mol, which was chosen to ensure an averaged temperature of 300K for the transition path ensemble. All the averaged quantities discussed in the text were averaged over 3,5000 trajectories. The committor for each configuration was estimated with 1000 shooting trajectories. For larger systems, computational cost for evaluating committors could be significantly reduced with a fitting procedure we recently developed [48].

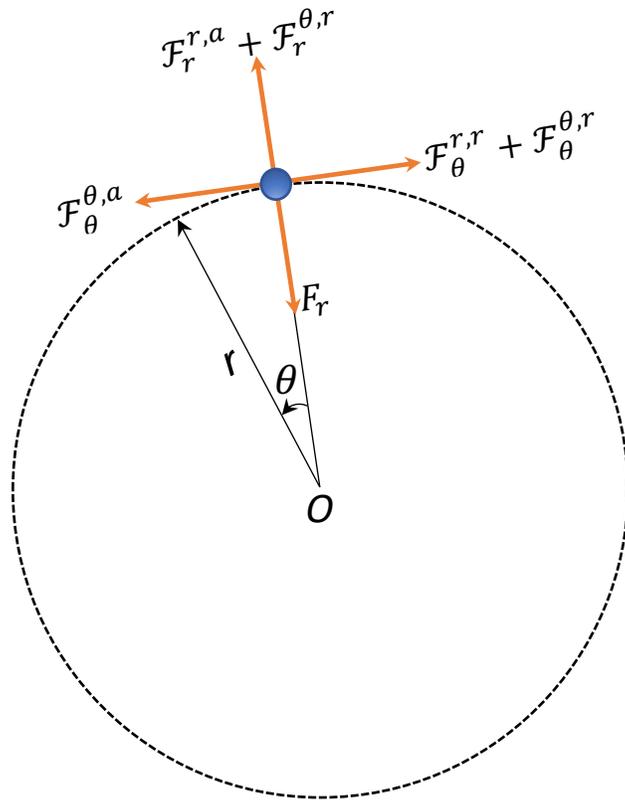

**Figure 1**: A schematic of a particle moving in a central force field, with the radial and angular coordinates, and the impressed and inertial forces labelled.

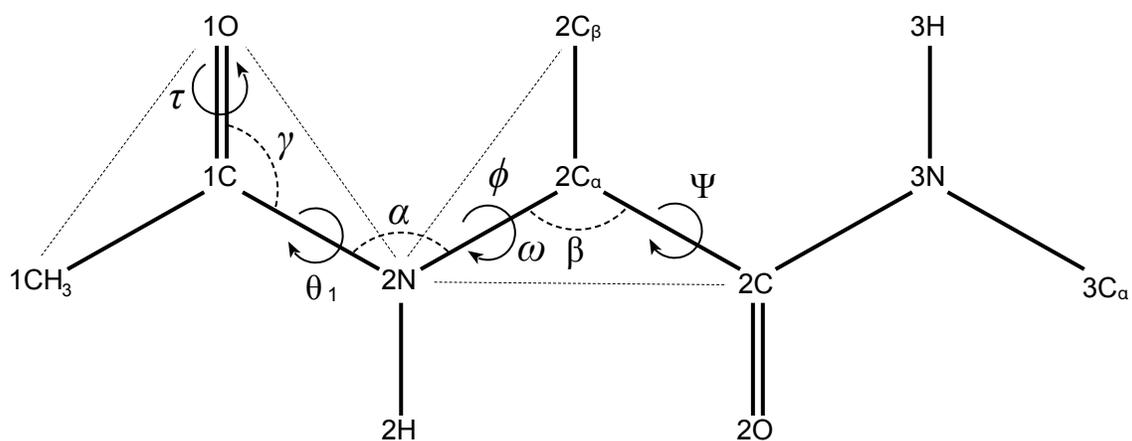

**Figure 2**: A schematic of the alanine dipeptide molecule with the coordinates discussed in the main text labelled. For proper dihedrals, we only mark the single bond that defines the relevant rotation. For improper dihedrals, each of the two planes that span the dihedral are defined by three atoms. We connect the two atoms that are not bonded to each other with a dotted line, so that it forms a triangle with the two bonds connecting these atoms with the central atom.

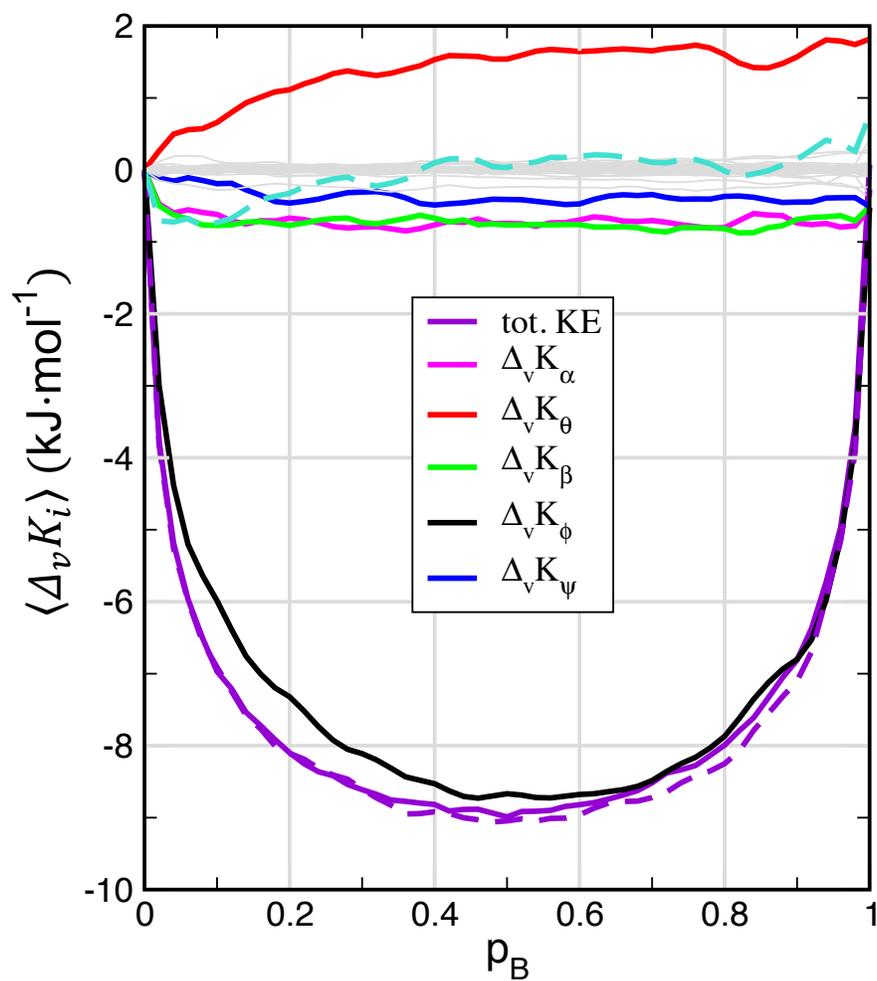

**Figure 3**: The KEFs through all the coordinates as a function of $p_B$. Violet dashed line: $\langle \Delta_v K_\phi \rangle + \langle \Delta_v K_{\theta_1} \rangle + \langle \Delta_v K_\psi \rangle + \langle \Delta_v K_\alpha \rangle + \langle \Delta_v K_\beta \rangle$. Turquoise dashed line: $\langle \Delta_v K_{\theta_1} \rangle + \langle \Delta_v K_\alpha \rangle + \langle \Delta_v K_\beta \rangle$. The Gray lines are KEFs through all the other 55 coordinates.

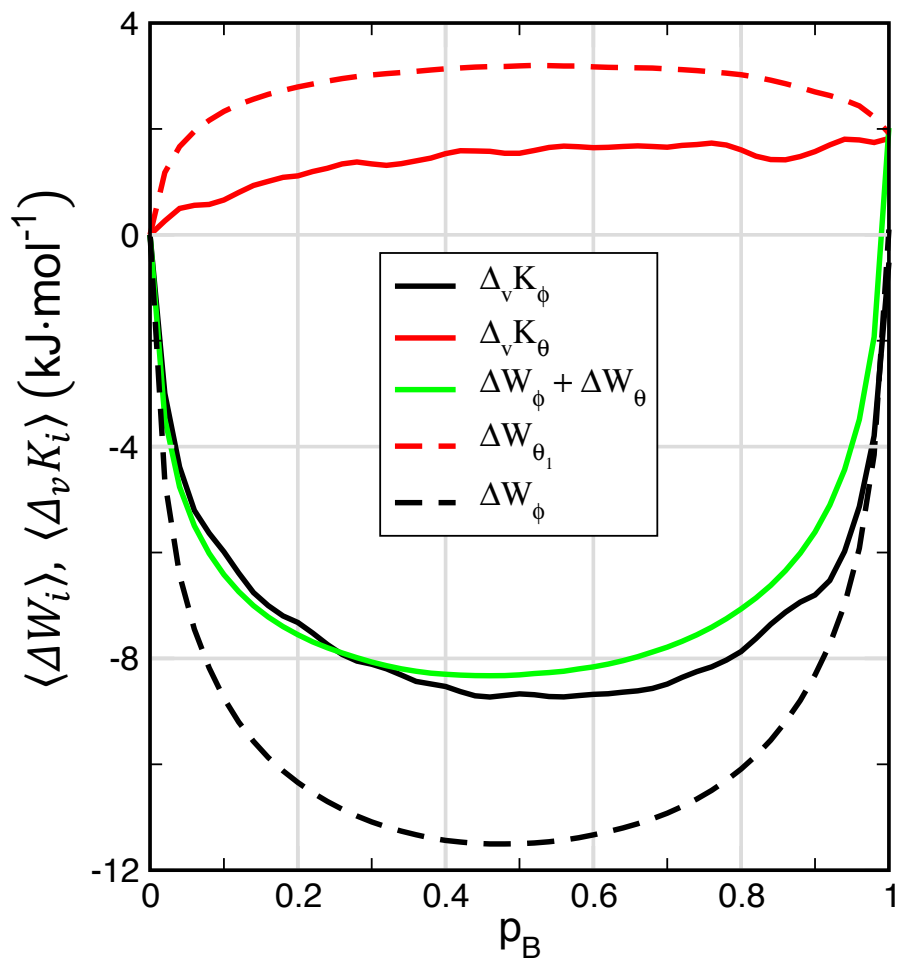

**Figure 4**: The PEFs and KEFs through $\phi$ and $\theta_1$.

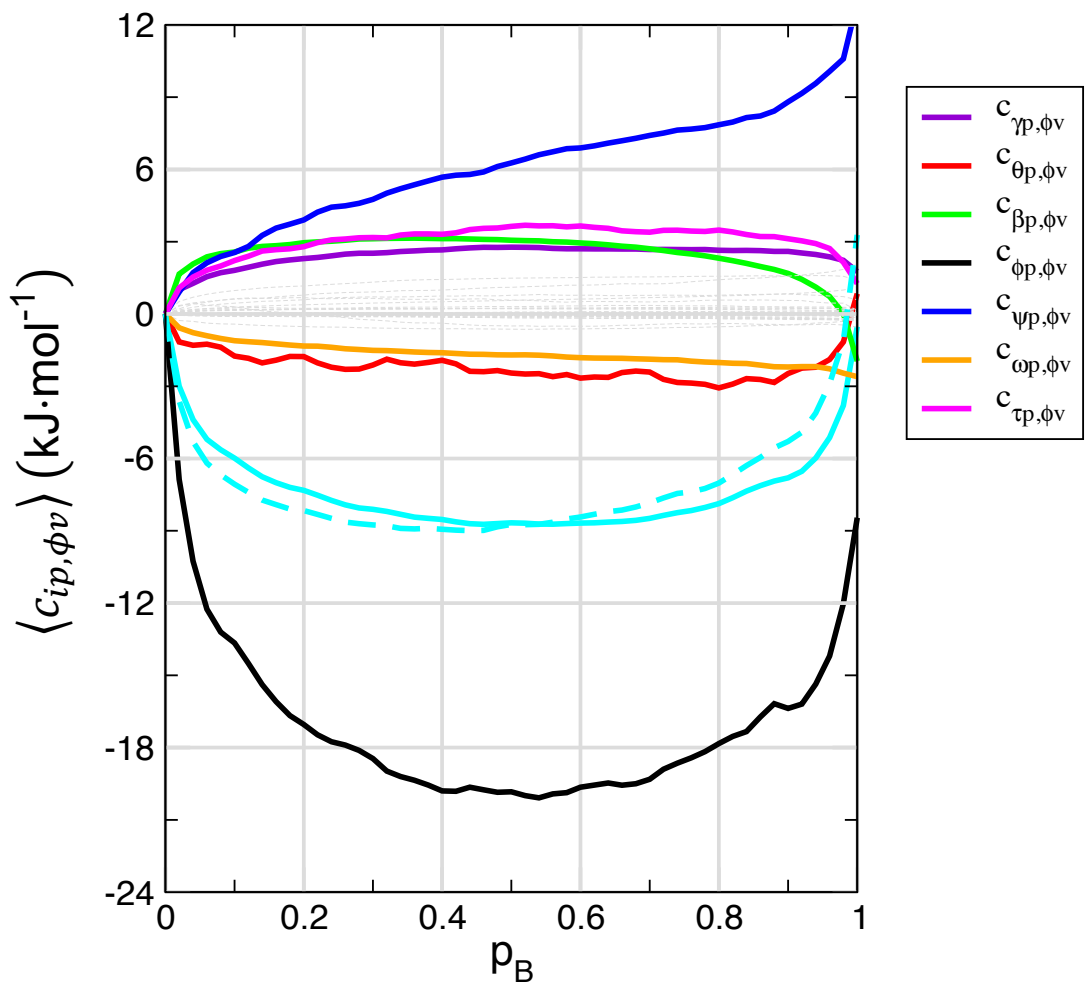

**Figure 5**: All the components of $\langle \Delta_v K_\phi \rangle$. Cyan solid line: $\langle \Delta_v K_\phi \rangle$. Cyan dashed line: $\langle c_{\phi p,\phi v} \rangle +$ $\langle c_{\theta_1 p,\phi v} \rangle + \langle c_{\psi p,\phi v} \rangle + \langle c_{\beta p,\phi v} \rangle + \langle c_{\gamma p,\phi v} \rangle + \langle c_{\omega p,\phi v} \rangle + \langle c_{\tau p,\phi v} \rangle$. Gray lines: $\langle c_{ip,\phi v} \rangle$ for all the other 53 coordinates in the system.

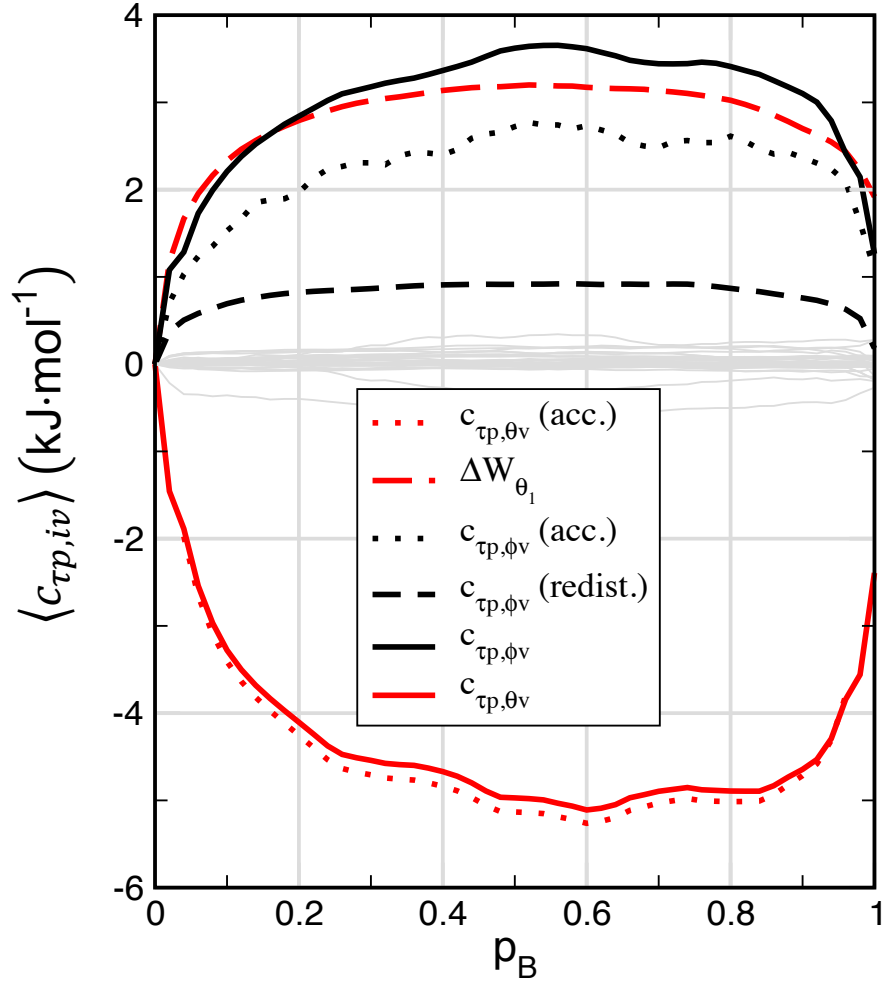

**Figure 6**: All the components of $\langle \Delta_p K_\tau \rangle$. Components of high magnitude are highlighted in color. For example, red dotted line denotes the component of $\langle c_{\tau p, \theta_1 v} \rangle$ due to the acceleration force from $\theta_1$. Gray lines are components that are vanishingly small.

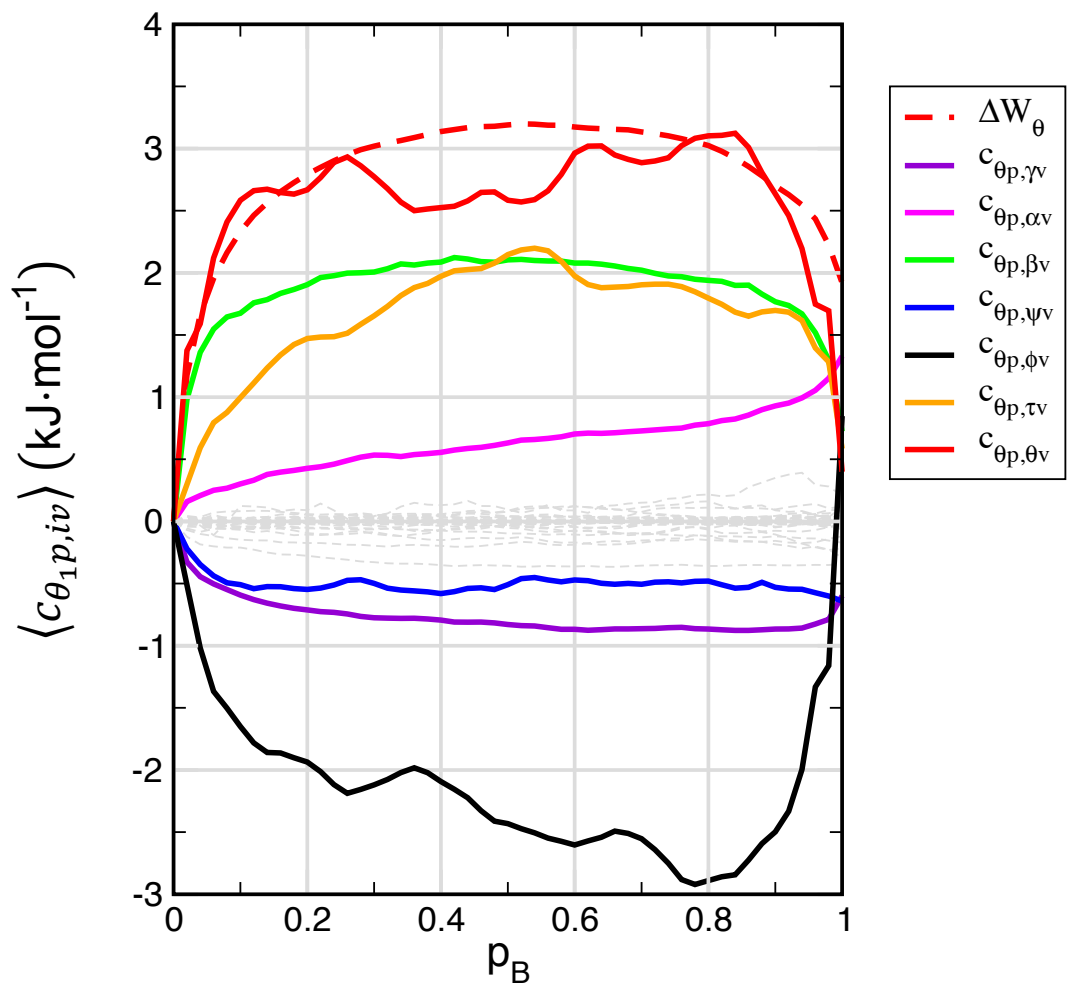

**Figure 7**: All the components of $\langle \Delta_p K_{\theta_1} \rangle$. Components of high magnitude are highlighted in color; gray lines are components that are vanishingly small. Red dashed line: $\langle \Delta W_{\theta_1} \rangle$.

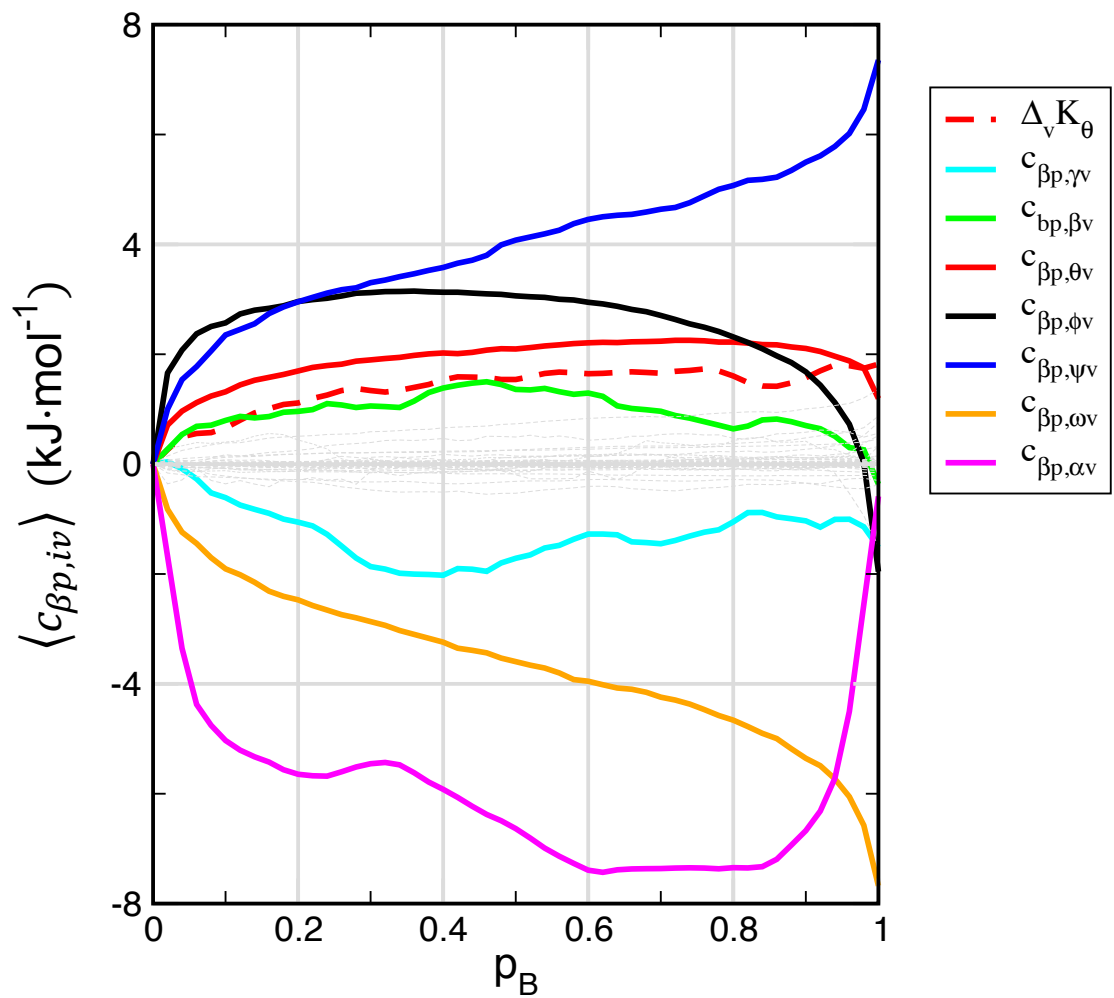

**Figure S1**: All the components of $\langle \Delta_p K_\beta \rangle$.

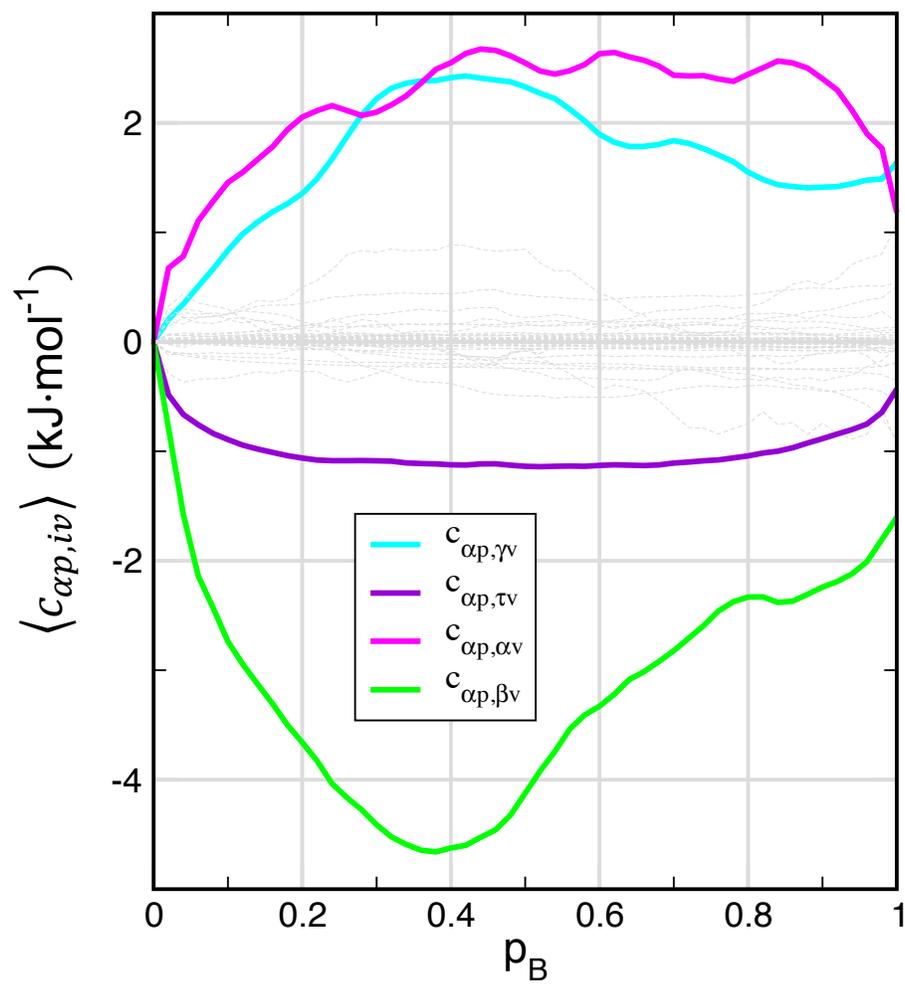

**Figure S2**: All the components of $\langle \Delta_p K_\alpha \rangle$.

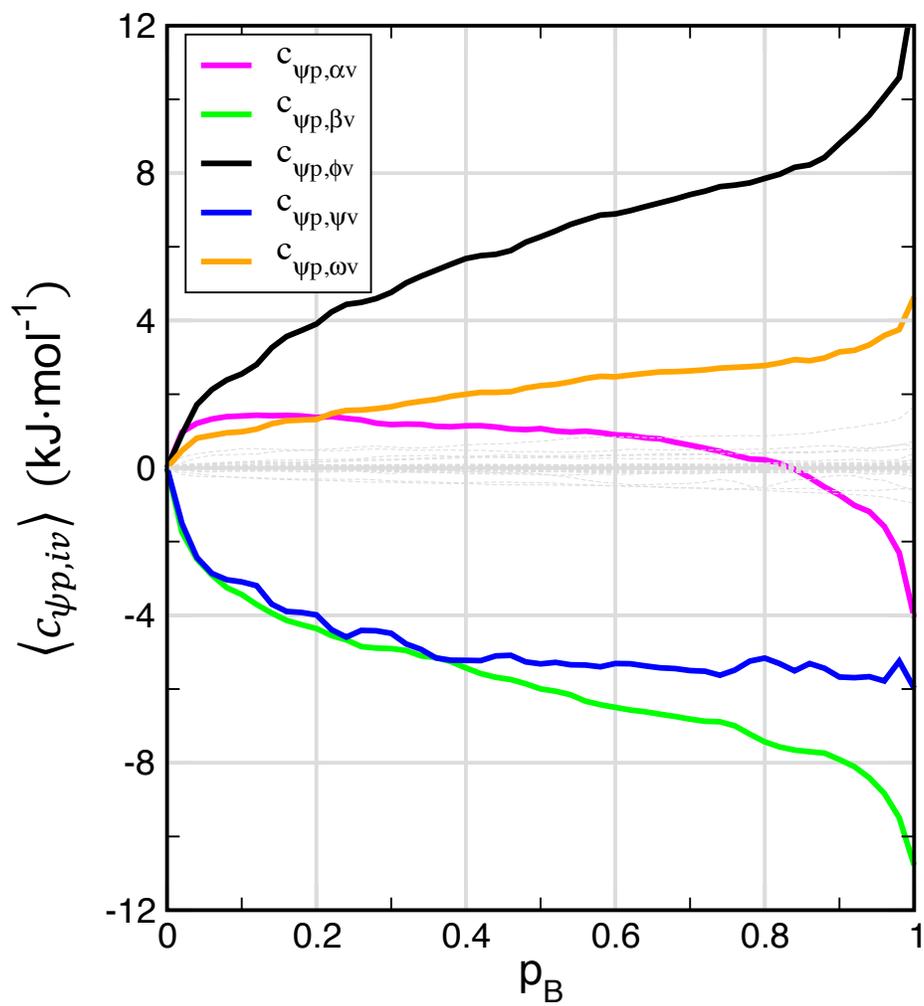

**Figure S3**: All the components of $\langle \Delta_p K_\psi \rangle$.

**Data Availability Statement**

The data that support the findings of this study are available from the corresponding author upon reasonable request.